\documentclass[preprint]{aastex}
\usepackage[]{epsfig}

\tightenlines
\newcommand{\Angstrom}	{\,{\rm \AA}}

\newcommand{\beq}	{\begin{equation}}
\newcommand{\cm}	{\,{\rm cm}}
\newcommand{\days}	{\,{\rm days}}

\newcommand{\eeq}	{\end{equation}}
\newcommand{\erg}	{\,{\rm ergs}}
\newcommand{\etal}	{{\it et al}}
\newcommand{\eV}	{\,{\rm eV}}
\newcommand{\fl}	{{\rm fl.}}

\newcommand{\gtsim}	{\gtrsim}		 %apj version
\renewcommand{\H}	{{\rm H}}
\newcommand{\He}	{{\rm He}}
\newcommand{\HH}	{{\rm H}_2}

\newcommand{\Jy}	{\,{\rm Jy}}

\newcommand{\kms}	{\,{\rm km~s}^{-1}}

\newcommand{\ltsim}	{\lesssim}		 %apj version
\newcommand{\Mpc}	{\,{\rm Mpc}}

\newcommand{\pc}	{\,{\rm pc}}
\newcommand{\s}		{\,{\rm s}}

\newcommand{\UV}	{{\rm uv}}

\begin{document}

\title{{\rm submitted to {\it Ap.\ J.\ Letters}\hfill POPe-807}\\
	\ \\
	H$_2$ Absorption and Fluorescence from 
	Gamma Ray Bursts in Molecular Clouds}

\author{B.T. Draine}
\affil{Princeton University Observatory, Peyton Hall, Princeton,
NJ 08544; draine@astro.princeton.edu}

\begin{abstract}
If a gamma ray burst with strong UV emission
occurs in a molecular cloud, there will be 
observable consequences
resulting from
excitation of the surrounding $\HH$.
The UV pulse from the GRB will pump $\HH$ into 
vibrationally-excited levels which produce strong absorption at
wavelengths $\lambda \ltsim 1650\Angstrom$.  As a result, both the
prompt flash and later afterglow will exhibit
strong absorption shortward of $1650\Angstrom$, with specific
spectroscopic features.
Such a cutoff in the emission from GRB\,980329 
may already have been observed
by Fruchter et al.; if so, GRB\,980329 was at redshift 
$3.0\ltsim z \ltsim 4.4$.
BVRI photometry of GRB\,990510 could also be explained by
$\HH$ absorption if GRB\,990510 is at redshift $1.6\ltsim z\ltsim 2.3$.
The fluorescence accompanying the UV pumping of the $\HH$ will
result in UV emission from the GRB which can extend
over days or months, depending on parameters
of the ambient medium and beaming of the GRB flash.
The 7.5--13.6~eV fluorescent
luminosity is $\sim 10^{41.7}\erg\s^{-1}$
for standard estimates of the parameters of the GRB and the ambient
medium.
Spectroscopy can distinguish this fluorescent 
emission from other possible sources of transient optical emission,
such as a supernova.
\end{abstract}


\section{Introduction}

At least some gamma ray bursts (GRBs) are accompanied by
intense optical emission, as demonstrated by detection of
a 9th magnitude optical transient coinciding with
GRB\,990123 (Akerlof \etal. 1999) and optical afterglows associated
with other GRBs (e.g., GRB\,990510: Stanek \etal. 1999; Israel \etal. 1999).
Since GRBs may be associated with
star-forming regions (Paczy\'nski 1998), it is of interest to consider 
observable phenomena which would be indicative of
molecular gas in the vicinity of the GRB.
Here we show that there may be both observable absorption and
emission due to $\HH$ within several parsecs of a GRB.

The energy radiated in the optical-UV flash and afterglow 
can be substantial, and
can have dramatic effects on interstellar
gas and dust in the vicinity of the GRB.
The $h\nu > 13.6\eV$ emission will photoionize the
gas, and
the pulse of optical radiation will vaporize
dust grains out to substantial distances from the GRB
(Waxman, Draine \& Phinney 1999).
The gamma-rays and hard X-rays emitted by the GRB will 
contribute to ionization of the nearby gas, but the UV and
soft X-rays have the dominant effect 
because of the much larger number of photons,
and much larger photoionization cross sections.

Ultraviolet radiation will destroy $\HH$
in the vicinity of the GRB, but before destruction the typical
$\HH$ molecule will be excited to a vibrationally excited state
by UV pumping.  While cold molecular hydrogen can
absorb only at wavelengths $\lambda < 1110\Angstrom$,
this vibrationally-excited $\HH$ 
can absorb strongly at wavelengths as long
as $1650\Angstrom$.
The characteristic wavelength-dependent absorption by this
vibrationally-excited $\HH$ should be imprinted on the
spectrum of the optical-UV flash when it reaches distant
observers.  Subsequent UV emission from the ``afterglow''
would also be subject to this
characteristic absorption, due to vibrationally-excited $\HH$
generated by the initial UV pulse plus additional 
UV from the afterglow.

In addition to imposing a characteristic absorption spectrum
on the optical transient, the optical-UV
flash will generate fluorescent UV emission from
surrounding $\HH$.  Because of light-travel time delays, 
the fluorescent emission will appear to extend over 
perhaps tens of days.
The emission is strong enough that it could be detectable,
and it would have a characteristic spectrum
which would distinguish it from other possible
sources of optical-UV emission.

\section{Photoexcitation of H$_2$}

Prior to the GRB, the surrounding $\HH$ is almost entirely in the
first two or three rotational levels ($J=0,1,2$) of the ground
vibrational level ($v=0$) of the ground electronic state (X$^1\Sigma_g^+$).
To photoexcite to the
B$^1\Sigma_u^+$ and
C$^1\Pi_u^{\pm}$ states requires $h\nu > 11.2$ and 12.3 eV, respectively.
To photoionize $\HH$ out of X$^1\Sigma_g^+(v=0,J=0)$ requires
$h\nu>15.4\eV$. 
The cold $\HH$ itself is therefore essentially transparent to $h\nu < 11.2\eV$
photons -- the dominant absorption for $h\nu < 11.2\eV$ 
is due to dust mixed with the $\HH$.

The first UV photons to arrive will most likely photoexcite 
the $\HH$ via Lyman or Werner
band transitions to the B or C electronic states, or photoionize
the $\HH$ to $\HH^+$.
$\HH$ which is photoexcited to the B or C states will decay
(in $\sim10^{-9}\s$) back to the ground electronic state, but 
typically to a vibrationally-excited level (e.g., $v=5$) 
with the rotational quantum number changed by $\Delta J=0,\pm2$.
The lifetimes of the vibrationally-excited levels ($\sim10^6\s$) are long 
compared to the timescale for photoexcitation or photoionization
[$\sim10^{-3}\s$ -- see eq.(\ref{eq:chi},\ref{eq:pion},\ref{eq:pexc})], 
so that
depopulation of the rovibrationally-excited levels of $\HH$ 
will be primarily by UV photoexcitation and photoionization.

If the luminosity of the GRB pulse in 7.5--13.6~eV photons
is $10^{49}L_{49}\erg\s^{-1}$, and
the spectrum $L_\nu\propto\nu^{-1/2}$ as expected from simple models
for emission from the forward shock
(see Waxman, Draine, \& Phinney 1999),
then at a distance $10^{19}R_{19}\cm$ 
from the GRB the specific energy density $u_\nu$ is given by
\beq
\nu u_\nu = 4\times10^{-14}\chi 
\left(1000\Angstrom/\lambda\right)^{1/2}\erg\cm^{-3},
\label{eq:u_nu}
\eeq
where
\beq
\chi = 1.1\times10^{13}L_{49}R_{19}^{-2}
\label{eq:chi}
\eeq
is the intensity at 1000\AA\ relative to the Habing (1968) estimate
for the local interstellar radiation field.

Stimulated emission in the UV transitions is negligible provided
$\chi\ll10^{19}$, so we consider photoexcitation out of level $i$ 
of X$^1\Sigma_g^+$ to vibration-rotation states of 
B$^1\Sigma_u^+$ and C$^1\Pi_u^{\pm}$, followed
by spontaneous decay to level $j$ of X$^1\Sigma_g^+$.
Let $T_{ji}$ be the rate for photopumping out of level $i$ into level $j$,
and let $\zeta_i^{(pd)}$ and $\zeta_i^{(pi)}$ be the rates for 
photodissociation and photoionization out of level $i$.
Spontaneous decay via quadrupole transitions can be neglected if
$\chi\gg10^4$, and collisional deexcitation can be neglected if
$n_\H/\cm^{-3}\ll\chi$.
If
\beq
T_{ii}\equiv
-\left[\sum_{j\neq i} T_{ji} + \zeta_i^{(pd)} + \zeta_i^{(pi)}\right] ~,
\eeq
then the $\HH$ abundances $p_i\equiv 2n(\HH(v_i,J_i))/n_\H$
evolve according to
\beq
\frac{d}{dt} p_i = \sum_j T_{ij}p_j ~.
\eeq
We consider the 299 bound states of $\HH$ with $J\leq 29$, and
construct photopumping rates as described by Draine \& Bertoldi (1996),
using Lyman and Werner band oscillator strengths and dissociation
probabilities from Abgrall \etal. (1993a,b) and Roueff (1993).
We neglect attenuation of the radiation field due to absorption by
$\HH$, H, or dust.
The rate of absorption of Lyman or Werner band photons for
$\HH$ in the $v=1, J=3$ level (for example), is 
\beq
\zeta_j^{(pexc.)}(v=1,J=3)=3.34\times10^{-10}\chi\s^{-1} ~,
\label{eq:pexc}
\eeq
with a probability $p_d\approx0.15$ of photodissociation.

If Lyman continuum radiation is present, then
the
$\HH$ photoionization cross section 
from Yan, Sadeghour, \& Dalgarno (1998) yields a photoionization rate
\beq
\zeta_j^{(pi)} = 1.96\times10^{-10}\chi\s^{-1},
\label{eq:pion}
\eeq
which we take to be independent
of rovibrational excitation.
We will also consider the case where all the Lyman continuum
has been absorbed, so that $\zeta_j^{(pi)}= 0$.

For the photoionization rate of (\ref{eq:pion}),
the probability of returning to a bound rovibrational state following
a photon absorption event is
$\alpha=(1-p_d)\zeta_j^{(pexc.)}/(\zeta_j^{(pexc.)}+\zeta_j^{(pi)})
\approx 0.54$, 
the probability of ultimately being destroyed by photodissociation (rather than
photoionization) is 
$p_d\zeta_j^{(pexc.)}/(p_d\zeta_j^{(pexc.)}+\zeta_j^{(pi)})\approx0.20$,
and the mean number of
fluorescent photons emitted per $\HH$ destroyed by radiation is
\beq
N_\fl = \frac{1}{(1-\alpha)} - 
\frac{\zeta_j^{(pi)}}{p_d\zeta_j^{(pexc.)}+\zeta_j^{(pi)}}
\approx 1.36 ~,
\label{eq:N_fl}
\eeq
including transitions
to the vibrational continuum of X$^1\Sigma_g^+$.
If, however, the illuminating spectrum is cut off at the Lyman limit,
then $\zeta_j^{(pi)}=0$, and $N_\fl\approx 6.7$ -- about 5 times as
many fluorescent photons as for the case where Lyman continuum is
present.

A detailed study of the $\HH$ photoexcitation would require modelling the
time-dependent spectrum of the radiation as it advances through
the cloud.
For a preliminary estimate of the rovibrational excitation we neglect
changes in the spectrum of the UV pulse, and simply assume the $\HH$
to be irradiated by the spectrum (\ref{eq:u_nu}), with and without
Lyman continuum radiation.
We assume the $\HH$ to be ``cold'' at $t=0$, with $p_j(0)=0.5$ 
for $(v,J)=(0,0)$ and (0,1), 
and we follow the evolution of the level populations
$p_j(t)$ in the presence of the radiation field (\ref{eq:u_nu}) until
negligible $\HH$ remains ($\sum p_j < 10^{-6}$).
We compute the time-integrated level populations
$P_j=\int_0^\infty p_j dt$.
In Figure 1 we show the normalized level populations 
$\phi_j\equiv P_j/\sum_j P_j$ 
versus level energy $E_j$.
We see that even with Lyman continuum radiation, 
the time-averaged level populations have $\sim50\%$
of the $\HH$ in vibrationally-excited levels; if photoionization is
suppressed, however, $\sim70\%$ of the $\HH$ is in levels $v \ge 1$.

\section{Photoabsorption by UV-pumped H$_2$}

Waxman, Draine \& Phinney (1999) 
discuss the destruction of dust by the GRB flash,
and estimate the dust destruction radius to be
$R_d\approx 2.5\times 10^{19}\cm$ for a $\sim$10 sec pulse with
luminosity $L_{49}\approx 0.55$,
and density $n_\H\ltsim 10^5\cm^{-3}$.
Since we will estimate below that the radial extent of UV-pumped
$\HH$ will be considerably smaller than $R_d$, it seems likely that
dust destruction will take place so rapidly (in the leading edge of
the GRB optical pulse) that we can neglect the
effects of dust absorption and scattering when considering the
excitation and destruction of the $\HH$.

Photoexcitation by the GRB optical pulse produces
rovibrationally excited $\HH$ with 
permitted absorption lines in the $912-1650\Angstrom$ region.
Figure \ref{fig:I/I_0:R=100} shows the absorption, smoothed to a resolution
$R=\lambda/\Delta\lambda=100$, produced by column densities
$N(\HH)=10^{16}$, $10^{18}$, and $10^{20}\cm^{-2}$
with the distributions $\phi(v,J)$ from
Figure \ref{fig:poph2}.
We assume
Voigt profiles, approximated following Rodgers \& Williams (1974), with
Doppler broadening parameter $b=3\kms$,
for the 28765 permitted Lyman and
Werner band lines between levels with $J\leq29$.
It is clear that the rovibrationally-excited $\HH$ absorbs quite strongly
for $\lambda \ltsim 1650\Angstrom$ ($h\nu \gtsim 7.5\eV$): for
$N(\HH)\approx 10^{18}\cm^{-2}$, it can be seen from Figure 
\ref{fig:I/I_0:R=100} that
approximately 1/3 of the energy
between 1650 and 912\AA\ has gone into photoexcitation of $\HH$.
In the discussion below we will assume that all of the photons between
1650 and 912\AA\ are available for photoexcitation of $\HH$.

The energy radiated in the afterglow may be comparable to or even exceed
that emitted in a few tens of seconds around the emission peak.
Let $N_{\rm ion}$ be the total number of $h\nu>13.6\eV$ photons,
and $N_\UV$ be the total number of 7.5--13.6~eV photons.
For the $F_\nu\sim\nu^{-1/2}$ spectrum observed for the afterglow of
GRB\,990510 (Stanek \etal. 1999) and expected from simple models 
(e.g., Waxman, Draine, \& Phinney 1999), $N_\UV/N_{\rm ion}=0.35$.

Dust within $\sim2.5\times10^{19}\cm$ of the GRB will be vaporized by
$h\nu<7.5\eV$ photons in the initial flash, so dust absorption will be
neglected for the $h\nu>7.5\eV$ photons of interest here.  In an
infinite cloud, the GRB will fully ionize the hydrogen and helium
within a radius 
\beq R_{ion}\approx 1.5\times10^{19}
\left(\frac{E_{50}}{n_3}\right)^{1/3} \cm ~, 
\eeq
where $n_3 \equiv
n_\H/10^3\cm^{-3}$, the total energy in 7.5--13.6~eV photons is
$10^{50}E_{50}\erg$, and we have assumed
$n_\He/n_\H=0.1$.  
For $r < R_{\rm ion}$ we will suppose that 20\%
of the $\HH$ is destroyed by photodissociation rather than by
photoionization, (absorbing $N_\fl=1.36$ UV photons per
photodissociation) thus absorbing a fraction $0.2N_\fl(0.5N_{\rm
ion}/1.2)/N_\UV = 0.33$ of the 7.5--13.6~eV photons at $r<R_{\rm
ion}$.  The remaining 67\% of the UV photons will photodissociate
$\HH$, with a probability $p_d\approx0.15$ per absorption, and
therefore will dissociate the $\HH$ out to a radius 
\beq R_{\HH}
\approx R_{\rm ion} \left(1+\frac{0.15\times0.67 (N_\UV/0.5)} {N_{\rm
ion}/1.2}\right)^{1/3} = 1.027 R_{\rm ion} 
\eeq 
The column density of this photodissociation zone is
$n_\H (R_{\HH}-R_{\rm ion}) = 4\times 10^{20}n_3^{2/3}E_{50}^{1/3}\cm^{-2}$.

A more accurate
treatment must simultaneously model photodissociation of the $\HH$,
photoionization of H, $\HH$, and He, and attenuation of the radiation
field by these absorption processes.  This is particularly critical
since we estimate $R_{\HH}\approx R_{\rm ion}$.  For the present,
however, we can safely conclude that not all the $7.5-13.6\eV$ photons
which can pump $\HH$ will have been absorbed by the radius $R_{\rm
ion}$ where all of the $h\nu>13.6\eV$ photons are exhausted.
Therefore there will be a shell of rovibrationally excited $\HH$
outside $R_{\rm ion}$.  Since the linear extent of the zone of
partially-dissociated (and therefore rovibrationally excited) $\HH$ is
determined by the attenuation length for the 1650--912\AA\ photons
which can photodissociate the $\HH$, Fig. 2 would suggest that this
layer of partially-dissociated (and rovibrationally-excited) $\HH$
would have a column density $N(\HH)\approx 10^{20}\cm^{-2}$, and
therefore a linear extent $\Delta R \approx
2\times10^{17}n_3^{-1}\cm$.  Because the vibrationally-excited levels
have lifetimes $\gtsim10^6\s$ against spontaneous decay, and
collisional deexcitation is negligible for $n_3\ltsim 10$, the
vibrational excitation in the layer of partially-dissociated $\HH$
will persist for weeks after the GRB.

Attenuation by the rovibrationally-excited $\HH$ will produce a strong
``jump'' near $1650\Angstrom$
in the spectrum of both the optical flash and the afterglow.
The absorption is due to many distinct absorption lines, as is
evident if the spectrum is shown with a resolution $R=1000$, as in
Figure \ref{fig:I/I_0:R=1000}.

Fruchter (1999) found that 
the spectrum of the GRB\,980329 optical transient showed a strong drop
in flux between I (9000\AA) and R (6600\AA).
Fruchter interpreted this drop
as due to Lyman $\alpha$ ($\lambda=1215\Angstrom$)
absorption at redshift $z\approx5$, with the Lyman forest then
suppressing the flux at shorter wavelengths.
However, recent deep imaging and spectroscopy of the host galaxy
indicate that it must be at $z < 5$, and probably
$z < 4$ (Djorgovski \etal. 1999).
If GRB\,980329 occured in a molecular
region, the observed drop in flux between
I and R could have been 
due to the onset of $\HH$ absorption at 1650\AA\ in the host
galaxy,
in which case GRB\,980329 was at redshift
$3.0\ltsim z\ltsim4.4$.

Stanek \etal. (1999) reported that VRI photometry of the
afterglow of GRB\,990510 were described by $F_\nu\propto \nu^{-0.46\pm0.08}$,
but the B band flux falls below this prediction, indicating either a
deviation from power-law behavior in the source, or additional extinction
either near the GRB or in an intervening galaxy.
Here we note that rovibrationally-excited $\HH$ could account for this
if the redshift of GRB\,990510 is in the interval $1.6\ltsim z \ltsim 2.3$,
so that the $1650\Angstrom$ absorption ``edge'' falls between
B and V.
This redshift for GRB\,990510 is not inconsistent with the
detection of absorption lines at $z=1.62$ (Vreeswijk \etal. 1999a,b).

\section{H$_2$ Fluorescence}

As seen above, most of the $6.5-13.6 \eV$ photons emitted by a
GRB in a molecular cloud will be absorbed by $\HH$.  A fraction of the
photoexcitations will result in photodissociation, but approximately
80\% of the absorbed energy will be reradiated in UV fluorescent
emission.

Waxman, Draine \& Phinney (1999) 
have shown that the prompt optical flash will
result in destruction of the dust within $\sim 10\pc$ of the GRB.
If the observer is situated in the ``beam'' of the optical flash,
then the fluorescent emission can reach the observer along a path
from which the dust has been cleared,
out to a distance $R_d\approx10\pc$.
There will be some attenuation of the fluorescent emission by $\HH$
absorption, but such absorption is relatively weak 
for $\lambda \gtsim 1300\Angstrom$ (see Figure 2).

If the prompt optical flash is radiated into a cone with half-width
$\theta\approx0.1$ (for a beaming factor $4\pi/\pi\theta^2\approx
400(0.1/\theta)^2$),
then the fluorescent emission from a shell of radius $R$ will reach
the observer spread out over a time $\Delta t_{\HH \fl}\approx
R\theta^2/c\approx 39 (R/10^{19}\cm)(\theta/0.1)^2\days$.
Because $\Delta t_{\HH \fl}$ is long compared to the GRB flash, for
purposes of estimating the light curve of the fluorescent emission we can
approximate the flash itself as a delta function.

Suppose that a fraction $f(r)$ of the 
$\HH$ at radius $r$ is destroyed by the prompt flash,
with fluorescent energy $\epsilon$ emitted per destroyed $\HH$ molecule.
If the observer is situated along the beam axis, then the 
fluorescent emission from regions with time delay $< t$ is
\beq
E_\nu(t) = \pi \epsilon_\nu \theta^2 \int_0^{ct/\theta^2} f(r) n(\HH) r^2 dr
+ \pi \epsilon ct \int_{ct/\theta^2}^\infty f(r) n(\HH) rdr
\eeq
If we now assume that $n(\HH)$ is independent of radius, and
$f(r)=1$ for $r<R_{\HH}$, and 0 for $r>R_{\HH}$, then
the apparent luminosity is
\beq
(\nu L_\nu)_{\HH \fl} = \frac{dE_\nu}{dt} = 
\frac{\pi}{2} \nu\epsilon_\nu n(\HH) R_{\HH}^2 c
\left( 1 - \frac{c^2 t^2}{R_{\HH}^2 \theta^4} \right)
\eeq
\beq
= 
2.4\times 10^{41} 
\left(\frac{\nu\epsilon_\nu}{10^{-10}\erg}\right)
n_3
\left(\frac{R_{\HH}}{10^{19}\cm}\right)^2
\left\{ 1 - 
\left[
\left (\frac{t}{39\days}\right) 
\left(\frac{10^{19}\cm}{R_{\HH}}\right)
\left(\frac{0.1}{\theta}\right)^2
\right]^2
\right\}
\erg\s^{-1} ~.
\label{eq:Lfluor}
\eeq

We have computed the fluorescence spectrum $\nu\epsilon_\nu$ for $\HH$ being
destroyed by continuum radiation with $I_\nu\propto\nu^{-1/2}$,
both with and without a cutoff at 912$\Angstrom$; the
results are shown in Figure \ref{fig:emission}.
The emitted radiation is concentrated in the $1650-912\Angstrom$ (7.5-13.6 eV)
range.  
The values of $\nu\epsilon_\nu$ differ by a factor $\sim$5, but according
to our estimate, $\sim$90\% of the destroyed $\H$ is in the
photoionized zone, so $\nu\epsilon_\nu\approx 1\times10^{-10}\erg/\HH$
is a representative average.
There will be some ``self-absorption'' of this emission by the 
rovibrationally-excited $\HH$ itself, an effect not included in the
present analysis.
If this fluorescent emission has also to traverse cold $\HH$, it will
suffer additional absorption, 
but only in a small number of absorption lines between
1110 and 912 $\Angstrom$.

The fluorescent emission estimated in eq. (\ref{eq:Lfluor}) 
appears to be a necessary consequence of a GRB in a molecular region.
It is intriguing to note that the estimated 
luminosity and timescale are comparable to what would emerge from a
supernova, although the spectrum of course is very
dissimilar.

Since the emission is concentrated between $1650-912\Angstrom$, this
process could not account for the
emission plateau observed  
$\sim 20-30$ days after GRB\,980326 (Bloom \etal. 1999) unless the redshift 
$z\approx 3.5$ (so that the 1600--1000~\AA\ emission would be redshifted
to 7200--4500~\AA).
However, at 
$z=3.5$ 
the observed $F_\nu = 0.4\mu{\rm Jy}$ at R would require
$\nu L_\nu=2\times10^{44}\erg\s^{-1}$
(for $\Omega=0.3$, $H_0=70\kms\Mpc^{-1}$), or
$n_3^{1/2}E_{50} \approx 5000$,
much larger than anticipated for standard parameters
($n_3\approx 1$, $E_{50} \approx 1$).
Furthermore, the spectrum obtained by Bloom \etal. 
28 days after the GRB does not appear consistent with a redshifted
version of Figure 4.
For GRB\,980326, then, it does not appear likely that 
the observed emission plateau is due
to $\HH$ fluorescence; the $z\approx1$ supernova hypothesis 
proposed by Bloom \etal.
appears much more plausible.

However, $\HH$ fluorescence could perhaps be seen in other GRBs.
GRBs at $z\approx1.5$ -- so that the $1600\Angstrom$ emission peak
is redshifted to $\sim$4000\AA\ -- offer the best prospect for detection
of this fluorescent emission, although it would still be quite faint
for nominal parameters: 
$(F_\nu)_{\rm max}\approx 2.5\times 10^{-9}n_3
(R_{\HH}/10^{19}\cm)^2\Jy$ 
at 4000\AA.

\section{Discussion}

The GRB pulse will create a large amount of $\HH^+$ mixed with surviving $\HH$
in the partially-dissociated shell.
In these regions $\H_3^+$ will be formed on a timescale
$\sim10^6(10^3\cm^{-3}/n(\HH))\s$.
This $\H_3^+$ will promptly react to produce other species, e.g.,
$\H_3^+ + {\rm CO} \rightarrow {\rm HCO}^+ + \HH$.
The millimeter-wave lines of HCO$^+$ and other species might
be observable in absorption against the afterglow, although this would
probably require that the redshift first be determined by other methods
(e.g. the $\HH$ absorption lines discussed here).

Perna \& Loeb (1998), Ghisellini \etal. (1998), and
B\"ottcher \etal. (1998) have previously called
attention to time-varying atomic and ionic absorption lines and
X-ray absorption edges expected
for GRBs, and Ghisellini \etal. and B\"ottcher \etal. pointed
out that time-varying Fe K-$\alpha$ fluorescence might be observable
from GRBs.
Perna, Raymond \& Loeb
(1999) have pointed out that atomic and ionic emission lines might
be observable from the photoionized remnants left behind by GRBs in 
relatively nearby galaxies.

Observations of any of these phenomena would provide valuable clues
to the nature of GRBs and their environments.
If GRBs occur in molecular
regions, the relatively strong
$\HH$ absorption spectrum may offer the best observational prospects 
for GRBs at redshift $z\gtsim 1.3$ so that $\lambda\approx 1650\Angstrom$
is accessible to ground-based spectrographs.

\section{Summary}

The principal conclusions of this paper are:
\begin{itemize}
\item Ultraviolet emission from a GRB will produce
vibrationally-excited $\HH$ which will produce strong line absorption
for $1650\gtsim\lambda\gtsim912\Angstrom$.  This absorption signature
should be imprinted on the spectrum of the radiation reaching us from
GRBs in molecular regions.

\item Absorption by vibrationally-excited $\HH$ could be responsible
for the pronounced drop in flux between R and I for GRB\,980329
(Fruchter 1999),
if GRB\,980329 is at $3.0 \ltsim z \ltsim 4.4$.

\item Absorption by vibrationally-excited $\HH$ could account for
the drop in flux between V and B for GRB\,990510
(Stanek \etal. 1999), if
GRB\,990510 is at $1.6\ltsim z \ltsim 2.3$

\item Ultraviolet emission from a GRB will produce fluorescent
emission from whatever $\HH$ may be present within tens of parsecs of
the GRB.  This fluorescence is potentially observable [see
eq. (\ref{eq:Lfluor}) and Fig.\ \ref{fig:emission}].

\end{itemize}

\acknowledgements

I am grateful to Bohdan Paczy\'nski for many valuable discussions, including
pointing out the possible application
to GRB\,980329, 
to S.G. Djorgovski for communicating results on GRB\,980329 in advance of
publication,
to E. Roueff for making available $\HH$ wavelengths and oscillator strengths,
and
to Robert Lupton for availability of the SM software package.
This work was supported in part by
NSF grant AST-9619429.

%--------------------------------------------------------------------------

%--------------------------------------------------------------------------

\newcommand{\figwidth}{6.0in}
\begin{figure}
\epsfig{file=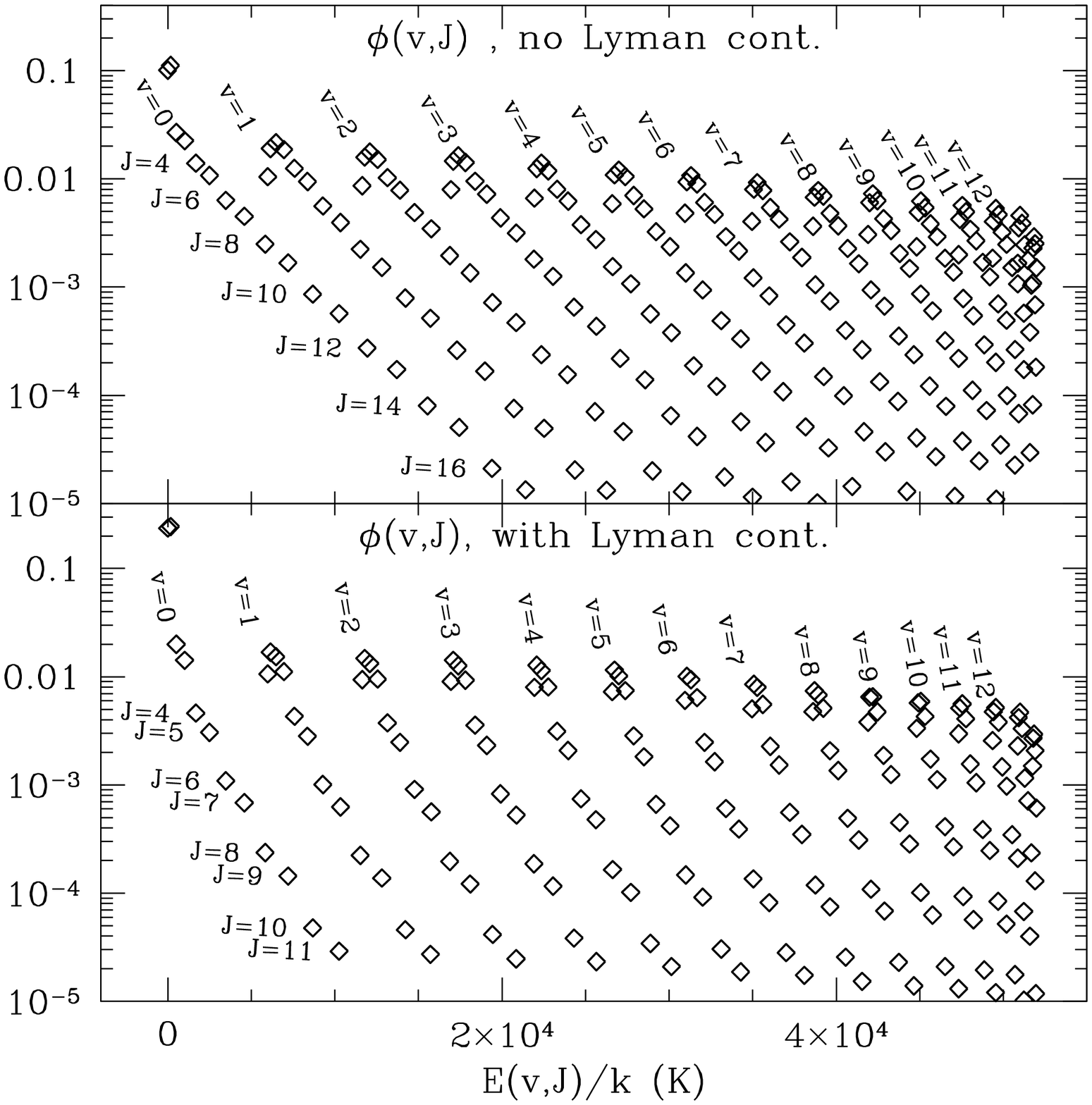,width=\figwidth}
\caption{
	\label{fig:poph2}
	Time-averaged fraction of the irradiated $\HH$ in different
	vibration-rotation states, versus energy of the level, for
	an irradiating spectrum $I_\nu\propto\nu^{-1/2}$, with
	intensity $10^5 \ll \chi \ll 10^{19}$.
	For a typical GRB flash 
	with $L_{49}\approx 1$ we
	estimate $\chi\approx 10^{13}$ [see eq. (\ref{eq:chi})].
	Collisional processes
	are neglected, valid for $n_\H \ltsim 10^{-3}\chi\cm^{-3}$.
	Upper panel shows results for irradiation with
	cutoff at 912\Angstrom; lower panel shows result if
	there is no break at the Lyman limit.
}
\end{figure}

\begin{figure}
\epsfig{file=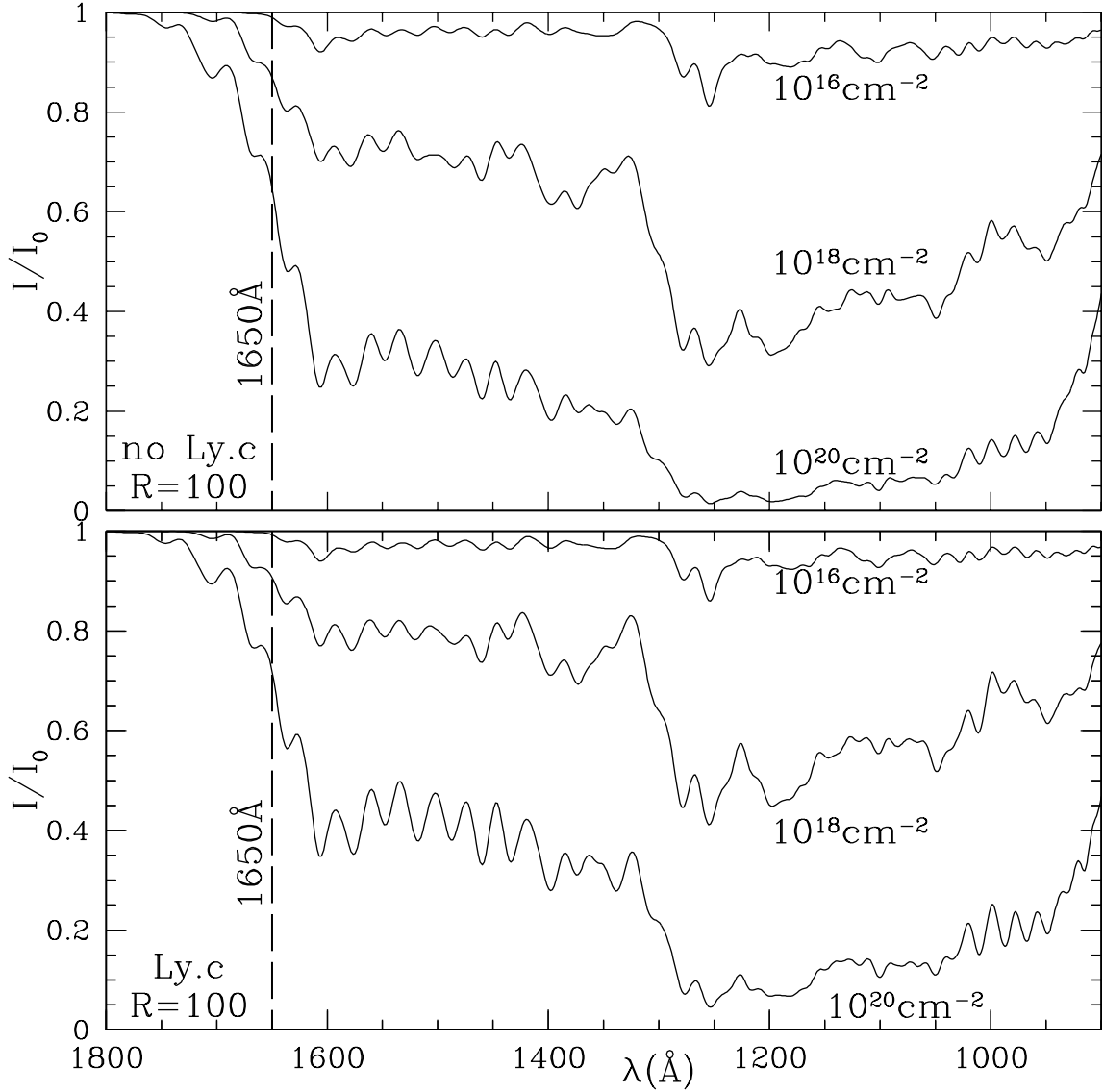,width=\figwidth}
\caption{
	\label{fig:I/I_0:R=100}
	Transmission through a medium with $\HH$ which has been
	rovibrationally excited by UV pumping by radiation with
	$I_\nu\propto\nu^{-1/2}$ and an intensity $10^5 \ll \chi \ll
	10^{19}$ relative to the Habing intensity.
	Transmission, smoothed to a resolution $R=100$, 
	is shown for column densities
	$N(\HH)=10^{16}$, $10^{18}$, and $10^{20}\cm^{-2}$ with
	the level populations of Fig. \ref{fig:poph2}.
}
\end{figure}

\begin{figure}
\epsfig{file=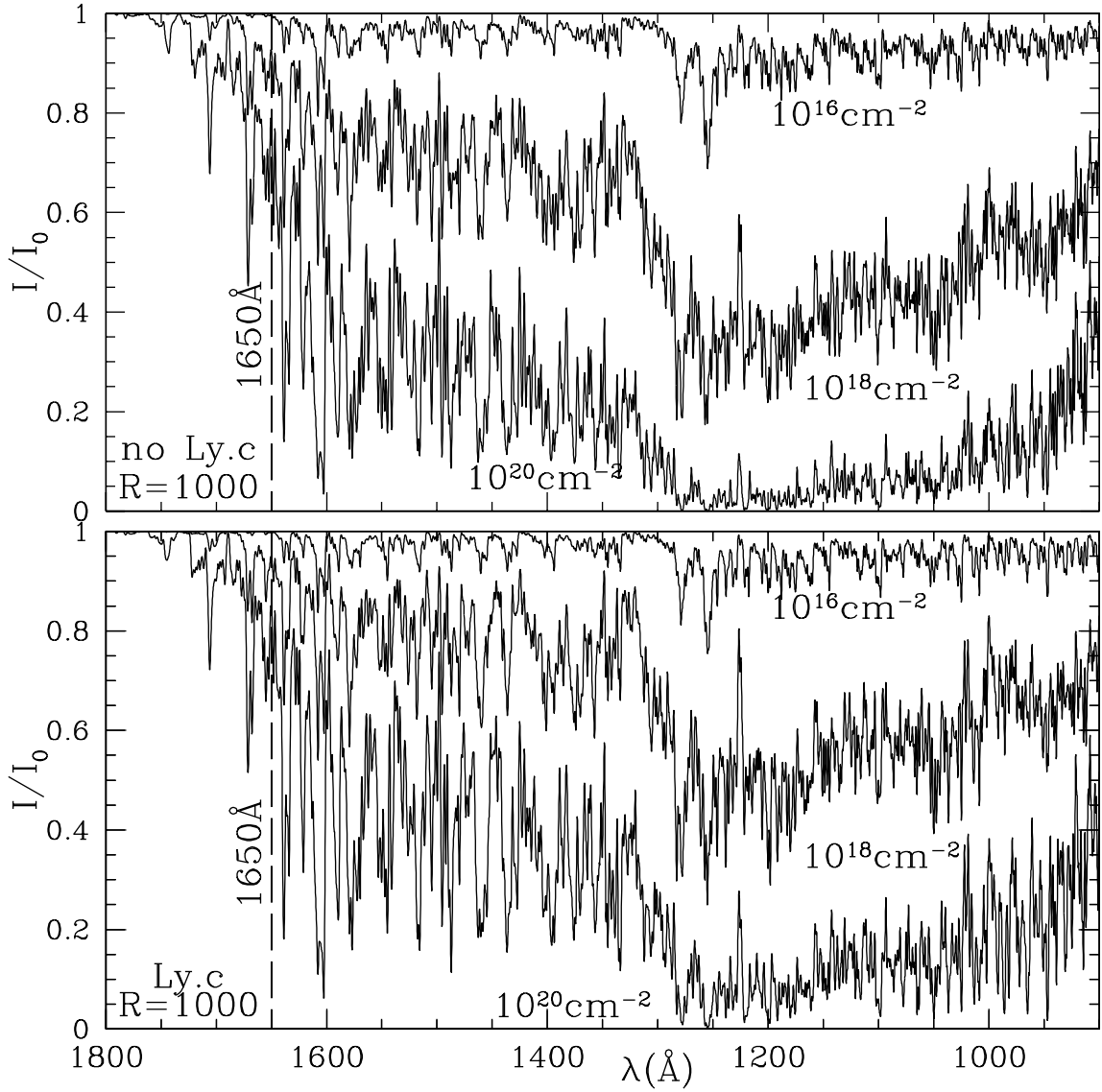,width=\figwidth}
\caption{
	\label{fig:I/I_0:R=1000}
	Same as Fig. \ref{fig:I/I_0:R=100}, but with resolution $R=1000$.
	}
\end{figure}
\begin{figure}
\epsfig{file=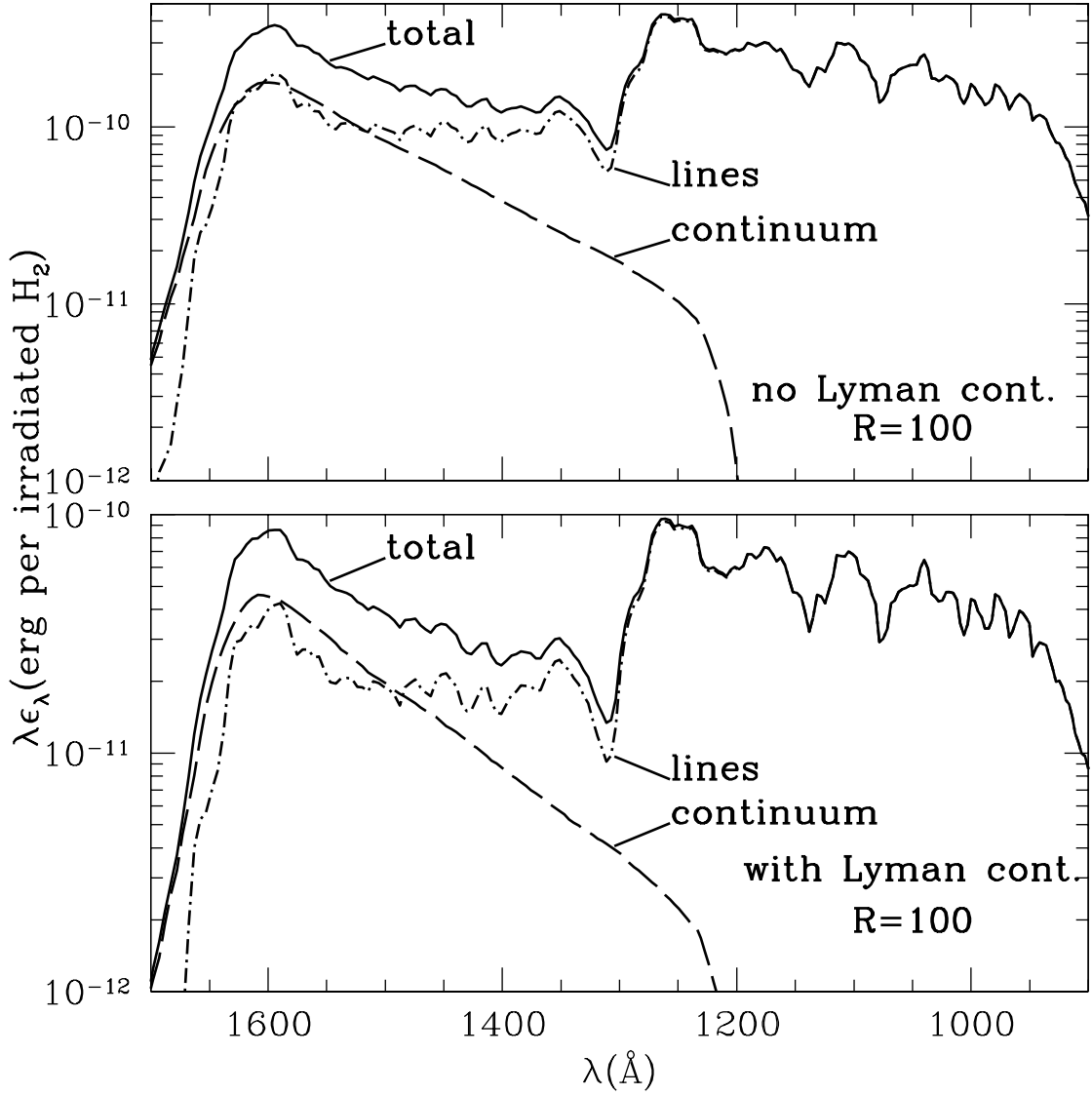,width=\figwidth}
\caption{
	\label{fig:emission}
	Fluorescent emission $\nu E_\nu$ per irradiated $\HH$
	molecule, for irradiation by $I_\nu\propto\nu^{-1/2}$
	with cutoff at $912\Angstrom$ (upper panel), or including
	Lyman continuum (lower panel).
	}
\end{figure}
\end{document}